\documentclass{article}
\usepackage{spconf,amsmath,graphicx}

\usepackage{amsfonts}
\usepackage{multirow}
\usepackage{hyperref}
\usepackage{subfigure}
\usepackage{booktabs,setspace}
\usepackage{threeparttable}
\ninept
\usepackage{setspace}

\usepackage{amssymb,CJK, indentfirst,balance,appendix}

\title{ROYALFLUSH SPEAKER DIARIZATION SYSTEM FOR icassp 2022 MULTI-CHANNEL MULTI-PARTY MEETING TRANSCRIPTION CHALLENGE}
\name{Jingguang Tian, Xinhui Hu, Xinkang Xu}
\address{Hithink RoyalFlush AI Research Institute, Zhejiang, China}

\begin{document}

%
\maketitle

\begin{abstract}
This paper describes the Royalflush speaker diarization system submitted to the Multi-channel Multi-party Meeting Transcription Challenge(M2MET). Our system comprises speech enhancement, overlapped speech detection, speaker embedding extraction, speaker clustering, speech separation and system fusion. In this system, we made three contributions. 
First, we propose an architecture of combining the multi-channel and U-Net-based models, aiming at utilizing the benefits of these two individual architectures, for far-field overlapped speech detection.
Second, in order to use overlapped speech detection model to help speaker diarization, a speech separation based overlapped speech handling approach, in which the speaker verification technique is further applied, is proposed. 
Third, we explore three speaker embedding methods, and obtained the state-of-the-art performance on the CNCeleb-E test set. 
With these proposals, our best individual system significantly reduces DER from 15.25\% to 6.40\%, and the fusion of four systems finally achieves a DER of 6.30\% on the far-field Alimeeting evaluation set.

\end{abstract}


\section{Introduction}
\noindent
Speaker diarization is a task to segment audio by speaker identity, or a task to determine “who spoke when”\cite{park2022review}. It is an important and valuable speech technology, especially in meeting scenarios, which can not only help improve speech recognition, but also to display speaker-annotated transcription results for better user experience. However, speaker diarization in real-world conferences presents many challenges due to overlapped speech, noises, and reverberation. 

A traditional speaker diarization system includes multiple modules: voice activity detection(VAD), speech segmentation, speaker embedding extraction and clustering. To tackle the problems caused by the adverse acoustic environments, some work has investigated the use of speech enhancement\cite{sun2018speaker}. Furthermore, speech separation (SS)\cite{xiao2021microsoft} and overlapped speech detection (OSD)\cite{landini2021analysis},  have been respectively shown to be effective in labeling speakers in overlapped regions. Recently, the end-to-end neural diarization systems and target-speaker voice activity detection (TS-VAD) demonstrated comparable or better performance to traditional methods in the CHIME-6\cite{medennikov2020target}, DIHARD \uppercase\expandafter{\romannumeral3}\cite{horiguchi2021hitachi} and VoxSRC 2021\cite{wang2021dku} challenges. However, the performance of these  schemes under unseen conditions is unclear and the number of recognizable speakers is limited by the output nodes of the neural network.

Multi-channel Multi-party Meeting Transcription(M2MeT) Challenge\cite{yu2021m2met} is a newly launched ICASSP2022 Signal Processing Grand Challenge. It consists of two tracks designed to improve speaker diarization and speech recognition for far-field multi-channel multi-speaker meeting speech. 
We participated the track of speaker diarization including two sub-tracks, sub-track \uppercase\expandafter{\romannumeral1} is restricted to  AliMeeting, Aishell-4\cite{2021AISHELL} and CNCeleb\cite{li2020cn} for system building while sub-track \uppercase\expandafter{\romannumeral2} allows participants to use any data. 
Here, we report our work on this task.

Our main contributions are as follows. %
First, we propose a novel multi-channel U-Net-based architecture for far-field OSD. Previous work has indicated that utilizing rich information of multi-channel signals is beneficial for both far-field speech recognition\cite{ganapathy20183} and far-field speaker recognition\cite{2019Multi}. We argue that far-field  OSD  can also benefit from multi-channel signals. The U-Net-based architecture first reached the state-of-the-art level in medical image segmentation\cite{ronneberger2015u} and achieved competitive performance in VAD\cite{wang2021dku2}, which is generally considered to be similar to OSD task. We combine the advantages of these two approaches for far-field OSD. 
Second, we introduce a method to use the OSD model to improve speaker diarization. 
Third, we explore three mainstream speaker embedding architectures, and obtained a state-of-the-art performance on the CNCeleb-E test set. 
We show that integrating these proposed methods can significantly reduce the diarization error rate (DER).

The structure of this paper is organized as follows. 
Section \ref{system description} presents our speaker diarization system. 
In section \ref{datapreparation}, we describe the dataset for system building. 
Section \ref{evaluation} reports our experiments and the evaluation results. 
The last section \ref{conclusions} concludes the paper.

\section{system description}
\label{system description}
\begin{figure}[t]
\includegraphics[scale=0.26]{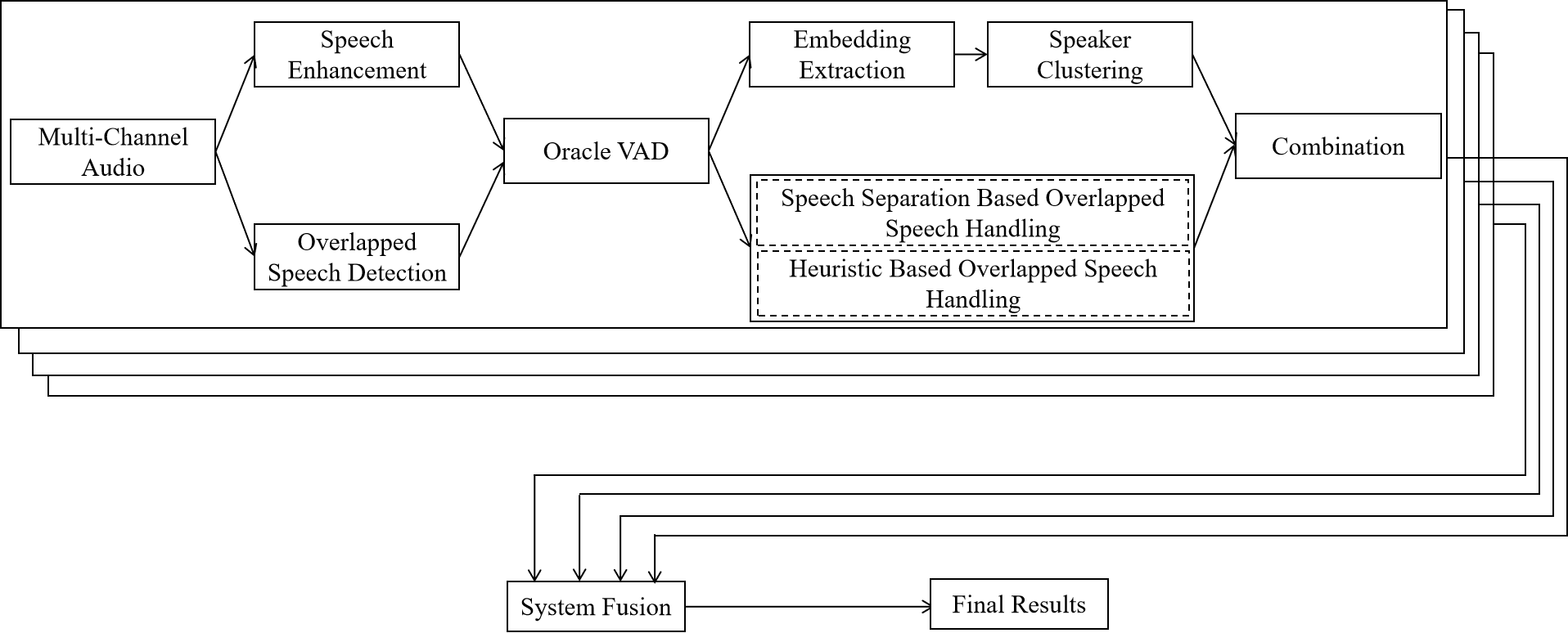}
\caption{System Diagram}
\label{system}
\end{figure}

\noindent
Our system is shown in Fig.\ref{system}, and it is a fusion of several sub-systems. Each sub-system is a conventional speaker diarization system, and it comprises speech enhancement, overlapped speech detection, speaker embedding extraction, speaker clustering and speech separation or heuristic based overlapped speech handling. 

The sub-system works like follows.
The 8-channel signal from the microphone array is directly input to an U-Net-based OSD model to judge if it is overlapped speech for each frame, and is converted to single-channel signal by speech enhancement module at the same time. 
Further processing is done on the single-channel signal. We use OSD model to divide oracle VAD segments into single speaker segments and overlapped speech segments. The single speaker segments are then split into smaller sub-segments, from each of which speaker embeddings are extracted. With the clustering module, we get speaker labels and speaker embedding centroids. The speech separation module processes overlapped speech segments. The speaker verification step is applied to separated speeches and speaker embedding centroids to obtain the speaker labels of overlapped speech segments. The speaker labels of single speaker segments and overlapped speech segments are put together as the system output. 
Finally, our final system is obtained by fusing multiple above sub-systems, using the DOVER-Lap \cite{raj2021dover} toolkit.

\begin{figure}[ht]
\includegraphics[scale=0.29]{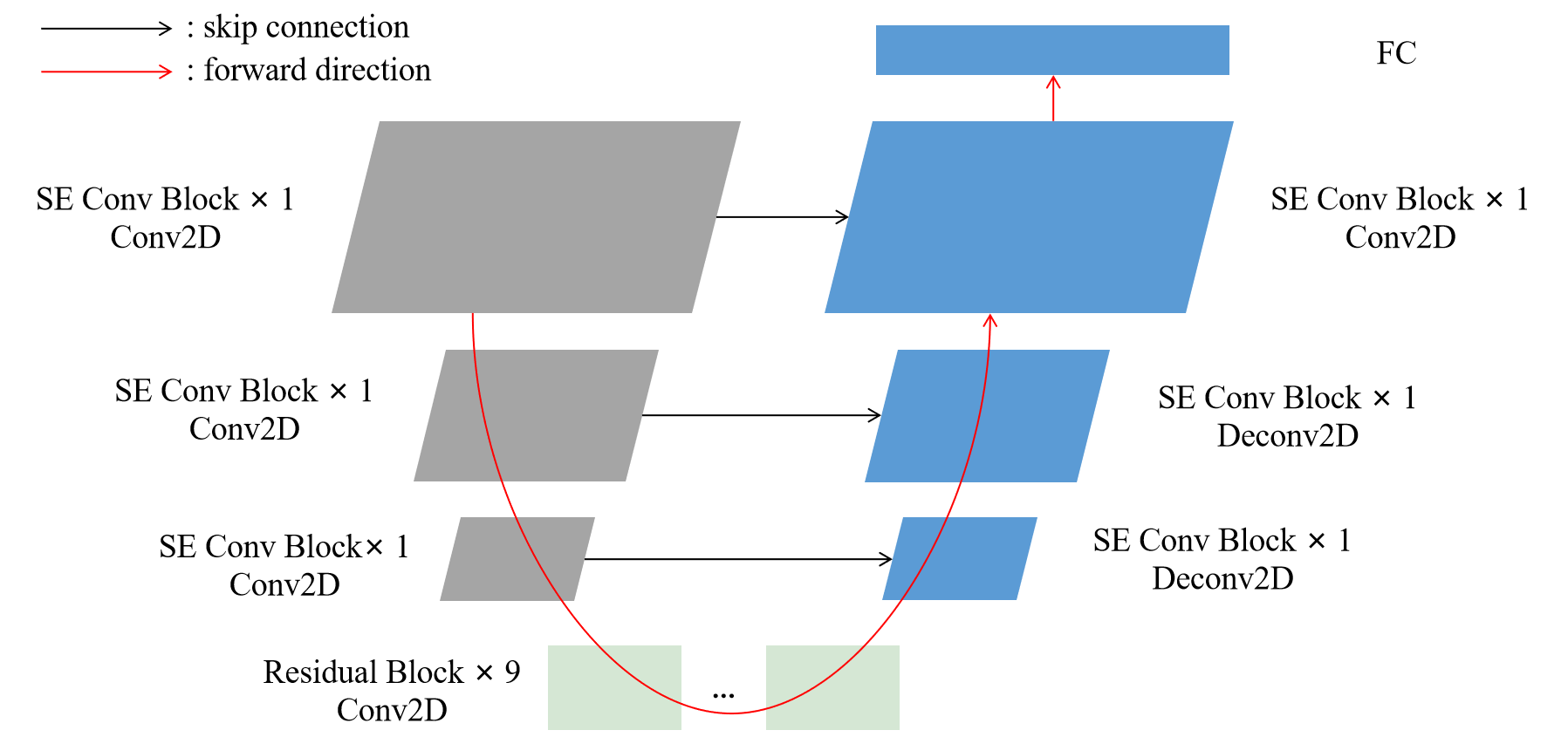}
\caption{the schematic of U-Net-based architecture}
\label{fig2}
\end{figure}

\subsection{Speech Enhancement}
\noindent
In this study, we applied two methods for speech enhancement. The weighted prediction error(WPE) is a signal dereverberation technique based on long-term linear prediction, which has been proven to be useful in far-field applications. First, the multi-channel WPE version implemented by NARA-WPE software package\cite{drude2018nara} is conducted. Then, after the WPE processing, the delay-and-sum(DAS) beamforming\cite{anguera2007acoustic}, which is a full acoustic beamforming frontend for denoising, is performed. 
Note that the original 8-channel signal is converted into a single enhanced signal after WPE and DAS beamforming. 
The subsequent processing is all done on the single-channel signal except for OSD.  In our experiments, we see that using multi-channel WPE is harmful for OSD while it is beneficial for speaker clustering and speech separation. Therefore, we do not apply any speech enhancement to OSD task.

\subsection{Overlapped Speech Detection}
\noindent
In this work, we explore two multi-channel training schemes based on the U-Net-based architecture. The input features are extracted independently from each channel of multi-channel audio. They can be viewed as a multi-channel 2D features or a 3D feature. The first scheme for multi-channel training is that the 2D features are directly input to a 2D convolutional neural network. The second scheme for multi-channel training is to feed 3D feature into a 3D convolutional neural network\cite{ji20123d}.

The schematic of multi-channel U-Net-based architecture is shown in Fig.\ref{fig2}, which is obtained by following \cite{johnson2016perceptual}. The U-Net-based architecture has three paths: the downsampling path, connection path and upsampling path. The downsampling path consists of three Squeeze-and-Excitation(SE) Conv blocks\cite{hu2018squeeze} involving Conv block and SE module. Conv block includes a convolutional layer, batch normalization(BN) and Rectified Linear Unit(ReLU). Note that the first SE Conv block does not downsample feature maps, the second and third downsample feature maps to half the size of the input, respectively. The connection path contains 9 repeated Residual blocks\cite{he2016deep} with constant feature map size. The upsampling path and downsampling path are symmetric and comprises three SE Conv blocks, where deconvolution is used to enlarge the feature map size. There are skip connections between the downsampling path and the upsampling path, and the corresponding feature maps are summed up as the input of the SE Conv block of the upsampling path. Finally, the feature maps generated by upsampling path are flattened along the channel and frequency dimensions, and fed to a fully connected(FC) layer. The detailed parameters of the architecture are shown in Table\ref{tab:table-U-Net}. 
\begin{table}[]
  \caption{The U-Net-based architecture, C(kernel size, stride) denotes the convolutional layer, the output size represents the channel, time, and frequency dimensions, respectively.}
  \label{tab:table-U-Net}
  \centering
  \begin{tabular}{ccc}
  \hline
  \textbf{Layer}     & \textbf{Structure} & \textbf{Output Size} \\ \hline
  SE Conv Block      & C(7 × 7, 1)        & 64 × 400 × 64        \\
  SE Conv Block      & C(3 × 3, 2)        & 128 × 200 × 32       \\
  SE Conv Block      & C(3 × 3, 2)        & 256 × 100 × 16       \\
  Residual Block × 9 & C(3 × 3, 1)        & 256 × 100 × 16       \\
  SE Conv Block      & C(3 × 3, 1/2)      & 128 × 200 × 32       \\
  SE Conv Block      & C(3 × 3, 1/2)      & 64 × 400 × 64        \\
  SE Conv Block      & C(7 × 7, 1)        & 64 × 400 × 64        \\
  FC                 & -                  & 128                    \\ \hline
\end{tabular}
\end{table}

We experiment with two types of U-Net-based architecture. 
The first type is shown as described above, and it is called SEUnet1. Multi-channel 2D features are used as the input to the SEUnet1. Since the 3D convolutional layers are used, the computational complexity of the neural network will be greatly increased. 
The second type of the U-Net-based architecture is a mix of 2D convolutional layers and 3D convolutional layers. We just added a new 3D convolutional layer to the first layer of the SEUnet1 architecture with a kernel size of 8 × 1 × 1 and a stride of 1, which we call SEUnet2.

The input features are 64-dimensional log Mel-filterbank energies with a frame length of 25ms and a frame shift of 10ms. Utterance-level Cepstral Mean Normalization (CMN) by subtracting the cepstral mean from the whole utterance is applied to all input features. The log Mel-filterbank energies from each channel of multi-channel audio are stacked together as input to the neural network. As a data augmentation step, we randomly mask the Mel-filterbank energies 0 to 10 bins in the frequency dimension. 

The models are trained to perform classification for three classes: silence, single speaker speech and overlap speech. The training sample length is 4 seconds(400 frames), and the mini-batch size during training is 32. The stochastic gradient descent (SGD) optimizer with weight decay 2e-5 and the softmax cross-entropy loss are employed. The initial learning rate is 0.01 and is multiplied by 0.9 after each epoch. 

We take the posterior probability of overlap speech node in the output layer to determine whether it is overlapped speech frame by frame. Inference is performed over length of 400 frames with a stride of 200 frames. The outputs on overlapped segments are averaged to obtain the final posterior probability. We set the threshold to 0.55, which is tuned on the far-field Alimeeting evaluation set. In order to reduce the false alarm of OSD, we refine the OSD results using oracle VAD labels.

For comparison, we also conduct experiments in a single-channel setting. We re-implemented the BLSTM architecture that is used for OSD as our baseline\cite{sajjan2018leveraging}. It has two layers of 128 bidirectional LSTM units with a projection layer, followed by an FC layer with 128 neurons. The results of SEUnet1 on single-channel speech input are also reported in this paper. All training configurations are the same as described above, except that the input is single-channel speech. 
\begin{table}[t]
\caption{The frame-level results of multi-channel OSD models.}
\label{tab:framelevelMulti}
\centering
\begin{tabular}{cccc}
\hline
\textbf{Model} & \textbf{Precision} & \textbf{Recall} & \textbf{F1} \\ \hline
SEUnet1        & 91.45\%            & 77.21\%         & 83.73\%     \\
SEUnet2        & 91.10\%            & 77.18\%         & 83.56\%     \\
fusion         & 92.23\%            & 77.19\%         & 84.04\%     \\ \hline
\end{tabular}
\end{table}

\begin{table}[]
\caption{The frame-level results of single-channel OSD models.}
\label{tab:frameresultsSingle}
\centering
\begin{tabular}{cccc}
\hline
\textbf{Model} & \textbf{Precision} & \textbf{Recall} & \textbf{F1} \\ \hline
BLSTM          & 88.15\%            & 64.03\%         & 74.18\%     \\
SEUnet1        & 90.37\%            & 69.34\%         & 78.47\%     \\ \hline
\end{tabular}
\end{table}

OSD model’s performance is using frame-level precision, recall and F1 score. Table \ref{tab:framelevelMulti} and Table \ref{tab:frameresultsSingle} shows the performance of the multi-channel models and single-channel models on the far-field Alimeeting evaluation set respectively. As shown in Table \ref{tab:frameresultsSingle}, the proposed SEUnet1 architecture brings an absolute F1 improvement of 4.29\% over the baseline, and as in Table \ref{tab:framelevelMulti}, the additional improvement using multi-channel training translates to an absolute F1 improvement of 5.26\%. Finally, the equal-weighted fusion of the posterior probabilities of SEUnet1 and SEUnet2 brings another 0.31\% absolute F1 improvement. Compared to baseline, F1 has significantly improved from 74.18\% to 84.04\%. In subsequent experiments, we use the fusion of SEUnet1 and SEUnet2 as the OSD results.

\subsection{Speaker Embedding Extraction}
\noindent
With the development of deep learning technology, the performance of speaker recognition systems has been greatly improved, and various network architectures have emerged. Here, we explore three representative speaker embedding architectures, including the TDNN\cite{snyder2018x}, the Resnet34\cite{2019BUT} and the ECAPA-TDNN\cite{desplanques2020ecapa}. 

The TDNN model that we used is based on x-vector topology. The original x-vector topology includes five TDNN layers that learn frame-level representations from acoustic features, a statistical pooling layer that combines mean and standard deviation of the frame-level representations and two FC layers. \cite{wang2021revisiting} has shown that the simple temporal standard deviation pooling(TSDP) can perform better than statistical pooling. Therefore, we replace the statistical pooling of the standard x-vector topology with TSDP, we called TDNN-TSDP in this paper. The speaker embedding is extracted from the first FC layer. 

The Resnet34 is another mainstream architecture in the speaker verification task, also called r-vector topology. It comprises of 34 stacked Residual blocks, with a statistical pooling layer and an FC layer. We use the TSDP version, which we call Resnet34-TSDP, in this paper. More details can be found in \cite{wang2021revisiting}.

The ECAPA-TDNN provides state-of-the-art speaker recognition performance through a careful architecture design. It introduces several enhancements methods based on the x-vector topology, including 1-dimensional SE-Res2Block, multi-layer aggregation, and channel-dependent attentive statistical pooling. We use a large version of the original paper for this model, with 1024 filters. More details can be found in \cite{desplanques2020ecapa}.

All speaker embedding models are trained using AAM-softmax\cite{deng2019arcface} with a margin of 0.15 and a scale of 32. The input features are 64-dimensional log Mel-filterbank energies processed with utterance-level CMN, the same as for training the OSD model. The training data is cut into 1.5-second length segments, and the mini-batch size for training is 128. The SGD optimizer with weight decay 2e-5 is employed. The initial learning rate is 0.1 and is multiplied by 0.9 after each epoch. 

Table \ref{tab:speakeremb} shows the equal error rate(EER) of the speaker embedding models on the CNCeleb-E test set, and the best one is the ECAPA-TDNN. The performance of ECAPA-TDNN in the third row is 8.82\% EER. As far as we know, it is the best performance publicly reported.
\begin{table}[]
\caption{the EER of the speaker embedding models.}
\label{tab:speakeremb}
\centering
\begin{tabular}{cc}
\hline
\textbf{System} & \textbf{EER} \\ \hline
TDNN-TSDP       & 13.74\%      \\
Resnet34-TSDP   & 9.32\%       \\
ECAPA-TDNN      & 8.82\%       \\ \hline
\end{tabular}
\end{table}

\begin{table*}[]
\caption{The DER of the diarization systems.}
\label{tab:DERresults}
\centering
\begin{tabular}{cccccc}
\hline
\textbf{System} & \textbf{Speech enhancement} & \textbf{Speaker embedding} & \textbf{Overlapped speech detection} & \textbf{Overlapped speech handling} & \textbf{DER} \\ \hline
baseline        & -                          & -                          & -                                 & -                                & 15.25\%      \\
1               & \checkmark                 & TDNN-TSDP                  & \checkmark                                 & heuristic                        & 7.74\%       \\
2               & \checkmark                 & Resnet34-TSDP              & \checkmark                                 & heuristic                        & 7.50\%       \\
3               & \checkmark                 & ECAPA-TDNN                 & \checkmark                                 & heuristic                        & 7.47\%       \\
4               & \checkmark                 & TDNN-TSDP                  & \checkmark                                 & speech separation based          & 6.79\%       \\
5               & \checkmark                 & Resnet34-TSDP              & \checkmark                                 &speech separation based          & 6.55\%       \\
6               & \checkmark                 & ECAPA-TDNN                 & \checkmark                                 & speech separation based          & 6.40\%       \\
7                                            & \multicolumn{4}{c}{fusing 3, 4, 5, and 6}                                                                      & 6.30\%              \\ \hline
\end{tabular}
\end{table*}

\subsection{Clustering}
\noindent
We use NMESC\cite{park2019auto} to group speech segments into clusters. Speaker embedding will be corrupted if overlapped speech is included. Therefore, overlapped regions are excluded from the subsequent processing. The single speaker segments are uniformly broken up into 1.5s sub-segments with a shift of 0.25s. The speaker embeddings are extracted on each sub-segments, and then clustered using NMESC. After that, the speaker labels of single speaker segments are obtained, and the speaker embedding centroids, which will be used in speech separation based overlapped speech handling, are computed by averaging speaker embeddings for each cluster. 

To determine the speaker number in the diarization process, we utilized the prior knowledge about the data set. Since the OSD model is not perfect, some overlapped segments will inevitably deteriorate the speaker embeddings, and lead to the wrong speaker number estimation of the clustering. In this work, we use the information that the maximum number of speakers is four in the Alimeeting corpus. The overlap ratio of each session is calculated by the OSD model. If the overlap ratio is greater than 20\%, we consider the number of speakers in the session to be four. The motivation to do that is based on the consideration that the higher the proportion of overlapped speech segments, the more speaker in the meeting.

\subsection{Speech Separation based Overlapped Speech Handling}
\subsubsection{Speech separation}
\noindent
We adopt the fully-convolutional time-domain audio separation network(Conv-Tasnet)\cite{luo2019conv}, which consists of encoder, separator and decoder, for speech separation. we choose the global layer normalization (gLN) method in the network. The training data is cut into segments with length of 1 second. The mini-batch size during training is 16. In this paper, we only consider the case where two speakers are speaking at the same time. More details can be found in \cite{luo2019conv}.

\begin{figure}[ht]
\includegraphics[scale=0.29]{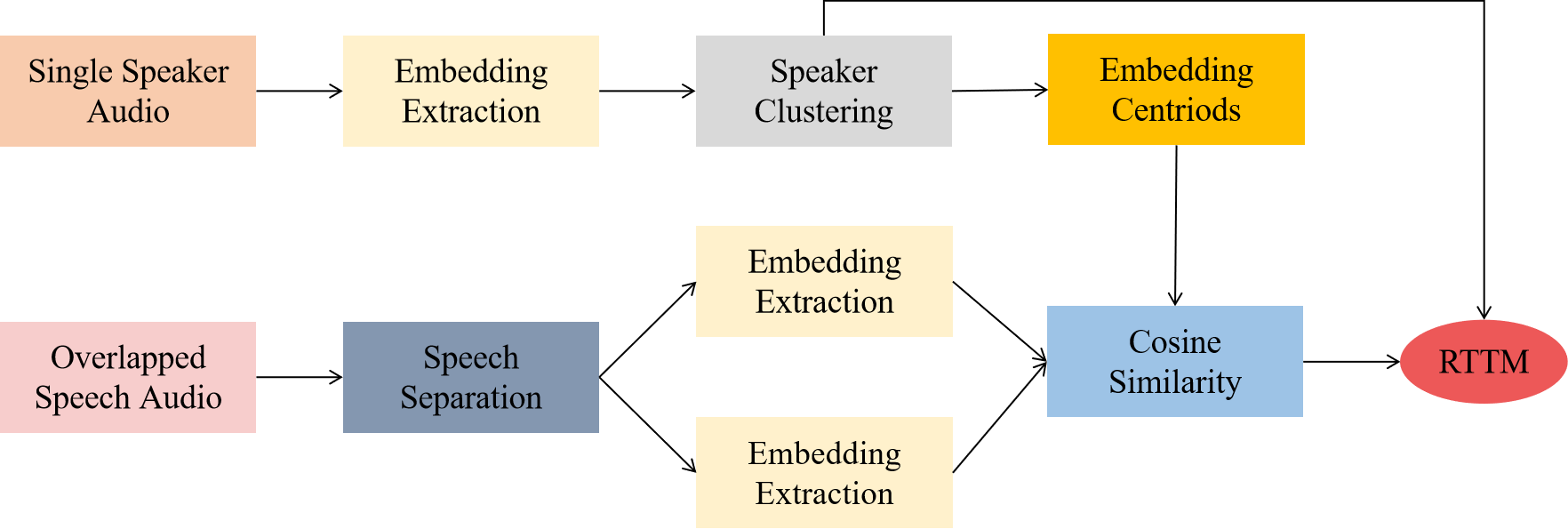}
\caption{speech separation based overlapped speech handling}
\label{fig-separation}
\end{figure}

\subsubsection{Overlapped speech handling}
\noindent
In this paper, we investigated two methods to handle the overlapped speech. We assume that the maximum number of speakers who can speak simultaneously is two. Once overlapped speech is detected, the two temporally closest speaker labels are assigned. We call this a heuristic method. In addition, we proposed another scheme as shown in the Fig. \ref{fig-separation}. In this scheme, the detected overlapped regions are input to the speech separation module, and then a speaker verification is processed. After these two steps, the separated speeches are assigned speaker labels. Specifically, for each overlapped speech segment, we take it as the input of Conv-Tasnet. The output has always two channels, each channel is supposed to contain only one speaker. The speaker embedding is extracted from both channels, respectively. Then, we allocate the two most likely speakers to overlapped speech segment according to the cosine similarity between speaker embedding centroids and speaker embedding of separated channels. Finally, The speaker labels of single speaker segments and overlapped speech segments are put together as the system output. 

\subsection{System Fusion}
\noindent
Since a speaker diarization task often benefits from an ensemble of several systems, we use DOVER-Lap to combine the outputs of different systems. The variations of our proposed systems are mainly reflected in the speaker embedding model and overlapped speech handling methods. In our final system that was used for submission of the challenge, we fused the outputs of the four systems shown in the Table \ref{tab:DERresults}, and will give further analyses later.

\section{Data preparation}
\label{datapreparation}
\noindent
The organizers of the M2MeT challenge released the AliMeeting corpus, consisting of 120 hours of real Mandarin meeting data recorded by 8-channel microphone array. 
\noindent
We follow the rules in sub-track \uppercase\expandafter{\romannumeral1} of the M2MeT. The detailed dataset used in this challenge for each model are as follows:

\begin{itemize}
\item \textbf{Overlapped speech detection:}
We used far-field Alimeeting train set(104 hours) and Aishell-4 train set(107 hours) to
train OSD models. The performance of models are reported in the far-field Alimeeting evaluation set.

\item \textbf{Speaker embedding:}
We only use the CNCeleb dataset having 2793 speakers as training data. The CNCeleb-E(which contains 200 speakers) is reserved as an evaluation set. We generate three extra samples for each utterance in the training data. The first two augmentations are generated using kaldi combining the publicly available MUSAN\cite{snyder2015musan} and the RIR\cite{ko2017study}. The remaining augmentation is generated by sox to tempo down or tempo up each utterance by 0.9 or 1.1. 

\item \textbf{Speech separation:}
 To train the speech separation model, we simulate 120 hours of mixed training data. The audio mixtures are generated by randomly selecting two clean utterances from different speakers in the far-field Alimeeting train set mixing at a random signal to noise ratio (SNR) between -5 $\sim$ 5 dB. 
\end{itemize}

\section{Results and Analysis}
\label{evaluation}
\noindent
The speaker diarization results of proposed systems and a baseline system released by the M2MeT office are shown in Table \ref{tab:DERresults}. Note that the official baseline system adopts a CHIME-6's VAD. For a fair performance comparison, we provide baseline results with oracle VAD in Table \ref{tab:DERresults}. 3 different speaker embedding model (the first set of sytems 1-3 or the second set of systems 4-6) are explored for experiments. Experimental results show that a more robust speaker embedding model results in a slightly better DER. Comparing the first set of systems 1-3 with the second set of systems 4-6, we can see that the DER of the overlapped speech handling based on speech separation, which is related to speaker verification,  is nearly 1\% better in absolute than the heuristic method. Comparing the baseline performance in the first row with our proposed systems, the best single system 6 reduced DER by 8.85\%  absolute. For submission of the challenge, systems 3, 4, 5 and 6 were fused using DOVER-Lap, and DER was further improved. Our final system achieved 6.30\% DER on the far-field Alimeeting evaluation set.

\section{Conclusions}
\label{conclusions}
\noindent
This paper described the Royalflush speaker diarization system submitted to the M2MeT Challenge. The main focuses are on multi-channel U-Net-based OSD, speaker embedding extractor, speaker clustering, speech separation based overlapped speech handling and system fusion. We evaluated the proposed system on far-field Alimeeting evaluation set, and significantly reduced DER from 15.25\% to 6.30\%. For future work, we will further introduce the End-to-End sub-system to the whole system.

\clearpage


\bibliographystyle{IEEEbib}
\bibliography{strings}

\end{document}